\documentclass[twocolumn]{aastex631}

\usepackage{natbib}
\usepackage{comment}
\usepackage{float}
\usepackage{wasysym}
\usepackage{placeins}
\usepackage{soul}
\usepackage{xcolor}

\newcommand{\hii}{$\mathrm{H}\,${\small II}}

\begin{document}

\title{Probing Shocked Ejecta in SN 1987A with XRISM$-$Resolve: the effects of the gate valve closed}

\author[0000-0002-6045-136X]{Vincenzo Sapienza}
\affiliation{Dipartimento di Fisica e Chimica E. Segr\`e, Universit\`a degli Studi di Palermo, Piazza del Parlamento 1, 90134, Palermo, Italy}
\affiliation{INAF-Osservatorio Astronomico di Palermo, Piazza del Parlamento 1, 90134, Palermo, Italy}
\affiliation{Department of Physics, Graduate School of Science, The University of Tokyo, 7-3-1 Hongo, Bunkyo-ku, Tokyo 113-0033, Japan}

\author[0000-0003-0876-8391]{Marco Miceli}
\affiliation{Dipartimento di Fisica e Chimica E. Segr\`e, Universit\`a degli Studi di Palermo, Piazza del Parlamento 1, 90134, Palermo, Italy}
\affiliation{INAF-Osservatorio Astronomico di Palermo, Piazza del Parlamento 1, 90134, Palermo, Italy}

\author[0000-0003-0890-4920]{Aya Bamba}
\affiliation{Department of Physics, Graduate School of Science, The University of Tokyo, 7-3-1 Hongo, Bunkyo-ku, Tokyo 113-0033, Japan}
\affiliation{Research Center for the Early Universe, School of Science, The University of Tokyo, 7-3-1 Hongo, Bunkyo-ku, Tokyo 113-0033, Japan}
\affiliation{Trans-Scale Quantum Science Institute, The University of Tokyo\\ 7-3-1 Hongo, Bunkyo-ku, Tokyo 113-0033, Japan}

\author[0000-0003-2836-540X]{Salvatore Orlando}
\affiliation{INAF-Osservatorio Astronomico di Palermo, Piazza del Parlamento 1, 90134, Palermo, Italy}

\author[0000-0002-2899-4241]{Shiu-Hang Lee}
\affiliation{Department of Astronomy, Kyoto University Oiwake-cho, Kitashirakawa, Sakyo-ku, Kyoto 606-8502, Japan}
\affiliation{Kavli Institute for the Physics and Mathematics of the Universe (WPI), The University of Tokyo, Kashiwa 277-8583, Japan}

\author[0000-0002-7025-284X]{Shigehiro Nagataki}
\affiliation{RIKEN Interdisciplinary Theoretical and Mathematical Sciences Program (iTHEMS), 2-1 Hirosawa, Wako, Saitama 351-0198, Japan}
\affiliation{ Astrophysical Big Bang Laboratory (ABBL), RIKEN Cluster for Pioneering Research, 2-1 Hirosawa, Wako, Saitama 351-0198, Japan}
\affiliation{Astrophysical Big Bang Group (ABBG), Okinawa Institute of Science and Technology Graduate University, 1919-1 Tancha, Onna-son, Kunigami-gun, Okinawa 904-0495, Japan}

\author[0000-0002-0603-918X]{Masaomi Ono}
\affiliation{RIKEN Interdisciplinary Theoretical and Mathematical Sciences Program (iTHEMS), 2-1 Hirosawa, Wako, Saitama 351-0198, Japan}
\affiliation{ Astrophysical Big Bang Laboratory (ABBL), RIKEN Cluster for Pioneering Research, 2-1 Hirosawa, Wako, Saitama 351-0198, Japan}
\affiliation{Institute of Astronomy and Astrophysics, Academia Sinica, Taipei 10617, Taiwan}

\author[0000-0002-1104-7205]{Satoru Katsuda}
\affiliation{Graduate School of Science and Engineering, Saitama University, 255 Simo-Ohkubo, Sakura-ku, Saitama city, Saitama, 338-8570, Japan}

\author{Koji Mori}
\affiliation{Faculty of Engineering, University of Miyazaki, 1-1 Gakuen Kibanadai Nishi, Miyazaki, Miyazaki 889-2192, Japan}
\affiliation{Japan Aerospace Exploration Agency, Institute of Space and Astronautical Science, 3-1-1 Yoshino-dai, Chuo-ku, Sagamihara, Kanagawa 252-5210, Japan}

\author[0000-0003-2008-6887]{Makoto Sawada}
\affiliation{Department of Physics, Rikkyo University, 3-34-1 Nishi Ikebukuro, Toshima-ku, Tokyo 171-8501, Japan}

\author[0000-0002-2359-1857]{Yukikatsu Terada}
\affiliation{Graduate School of Science and Engineering, Saitama University, 255 Simo-Ohkubo, Sakura-ku, Saitama city, Saitama, 338-8570, Japan}

\author[0000-0002-2774-3491]{Roberta Giuffrida}
\affiliation{Dipartimento di Fisica e Chimica E. Segr\`e, Universit\`a degli Studi di Palermo, Piazza del Parlamento 1, 90134, Palermo, Italy}
\affiliation{INAF-Osservatorio Astronomico di Palermo, Piazza del Parlamento 1, 90134, Palermo, Italy}

\author[0000-0002-2321-5616]{Fabrizio Bocchino}
\affiliation{INAF-Osservatorio Astronomico di Palermo, Piazza del Parlamento 1, 90134, Palermo, Italy}



\begin{abstract}
Supernova (SN) 1987A is widely regarded as an excellent candidate for leveraging the capabilities of the freshly launched XRISM satellite.
Recent researches indicate that the X-ray emission from SN 1987A will increasingly originate from its ejecta in the years to come.
In a previous study, we thoroughly examined the proficiency of XRISM-Resolve in identifying signatures of shocked ejecta in SN 1987A, synthesizing the XRISM-Resolve spectrum based on a state-of-the-art magneto-hydrodynamic simulation.
However, following the satellite's launch, a technical issue arose with the XRISM instrument's gate valve, which failed to open, thereby affecting observations with the Resolve spectrometer.
Here, we update our analysis, reevaluating our diagnostic approach under the assumption that the gate valve remains closed. We find that, even with the reduced instrumental capabilities, it will be possible to pinpoint the ejecta contribution through the study of the line profiles in the XRISM-Resolve spectrum of SN 1987A.
\end{abstract}

%%\keywords{Classical Novae (251) --- Ultraviolet astronomy(1736) --- History of astronomy(1868) --- Interdisciplinary astronomy(804)}


\section{Introduction} \label{sec:intro}

Supernova (SN) 1987A provides a unique opportunity to study the evolution of a SN into its remnant across the electromagnetic spectrum \citep{2016ARA&A..54...19M}.
The supernova remnant (SNR) is evolving in a clumpy equatorial ring within a diffuse \hii\ region \citep{2005ApJS..159...60S}.

X-ray observations are crucial for studying the interaction between the shock front, the circumstellar medium (CSM), and the ejecta.
Since the initial detection (\citealt{1987Natur.330..230D}; but see also \citealt{1994A&A...281L..45B}), the X-ray emission increased for about 25 years, confirming the shock's encounter with the equatorial ring (\citealt{1997ApJ...477..281B},\citealt{2005ApJ...634L..73P}, \citealt{2006A&A...460..811H}, \citealt{2009ApJ...692.1190Z}, \citealt{2012A&A...548L...3M}).
Recent investigations in the past few years, hinted that the X-ray emission from the CSM is decreasing, leaving space to the X-ray emission from the ejecta (\citealt{2016ApJ...829...40F}, \citealt{2021ApJ...916...41S}, \citealt{2021ApJ...922..140R}, \citealt{2022A&A...661A..30M}).

3D hydrodynamic and magneto-hydrodynamic simulations have replicated the X-ray light curves, spectra and remnant morphology, as well as other observed properties of SN 1987A.
Initially dominated by the shocked \hii\ region, the X-ray emission later became influenced by the shocked equatorial ring, and is now driven by the outer ejecta heated by the reverse shock \citep{2015ApJ...810..168O,2019A&A...622A..73O,2020A&A...636A..22O}.

A novel data analysis approach by \citet{2019NatAs...3..236M} using \textit{Chandra} data compared with 3D simulations provided insights into post-shock temperatures. 
The XRISM mission, launched on September 7, 2023, will offer high-resolution X-ray spectroscopy, allowing in-depth studies of SN 1987A. 
Using our in-house tool, we synthesized the XRISM-Resolve X-ray spectrum of SN 1987A from MHD simulations \citep{2020A&A...636A..22O,2020ApJ...888..111O}, successfully tracing new ejecta diagnostics for future observations \citep{2024ApJ...961L...9S}.

However, the XRISM Telescope safety gate valve for the Resolve instrument failed to open after launch.
As a consequence, the effective area of XRISM-Resolve results degraded and severely compromised under 2 keV.
Here we update the analysis in \citet{2024ApJ...961L...9S} by adopting the updated response files (Section \ref{sect:res}) and investigate the possibility of detecting the ejecta in SN 1987A with the gate valve closed (Section \ref{sect:con}).

%The structure of this research note is as follows:
%Section \ref{sect:res} showcases the results obtained from the synthesized models. 
%Discussions and conclusions are drawn in Section \ref{sect:con}.

\section{Results} \label{sect:res}
Following the same procedure described in \cite{2024ApJ...961L...9S}, we produced the synthetic spectrum of SN 1987A for the year 2024, using the new RMF and ARF files that take into account the effects of the gate valve closed, available in the HEASARC XRISM website\footnote{\url{https://heasarc.gsfc.nasa.gov/docs/xrism/proposals/responses.html}}.

\begin{figure*}[ht!b]
    \centering
    \includegraphics[width=\textwidth]{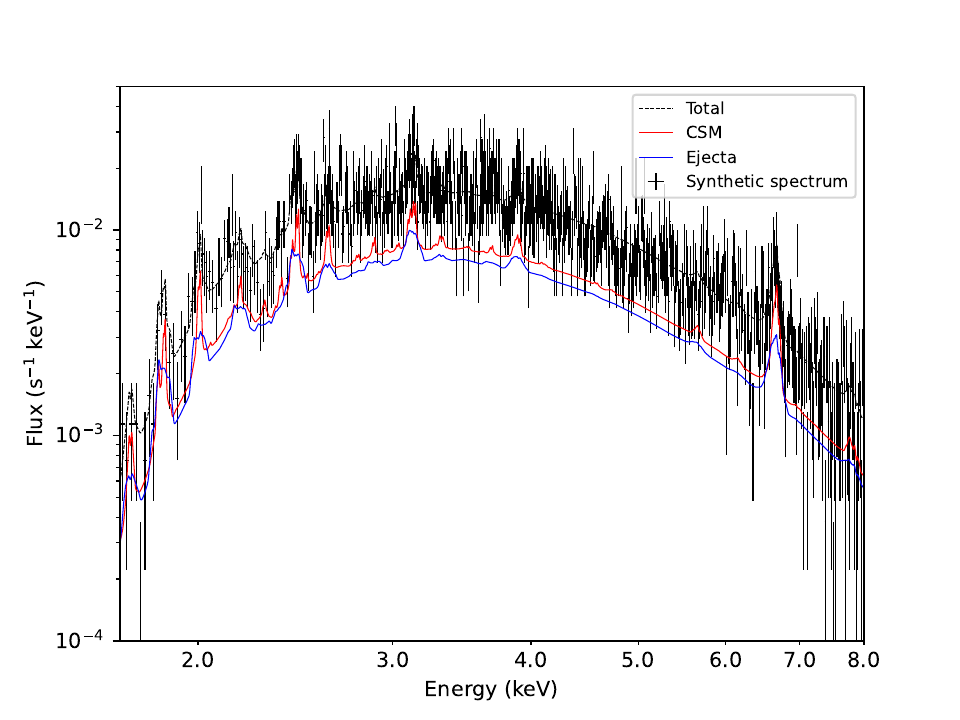}
    \includegraphics[trim={10pt 30pt 20pt 20pt},clip,width=0.97\columnwidth]{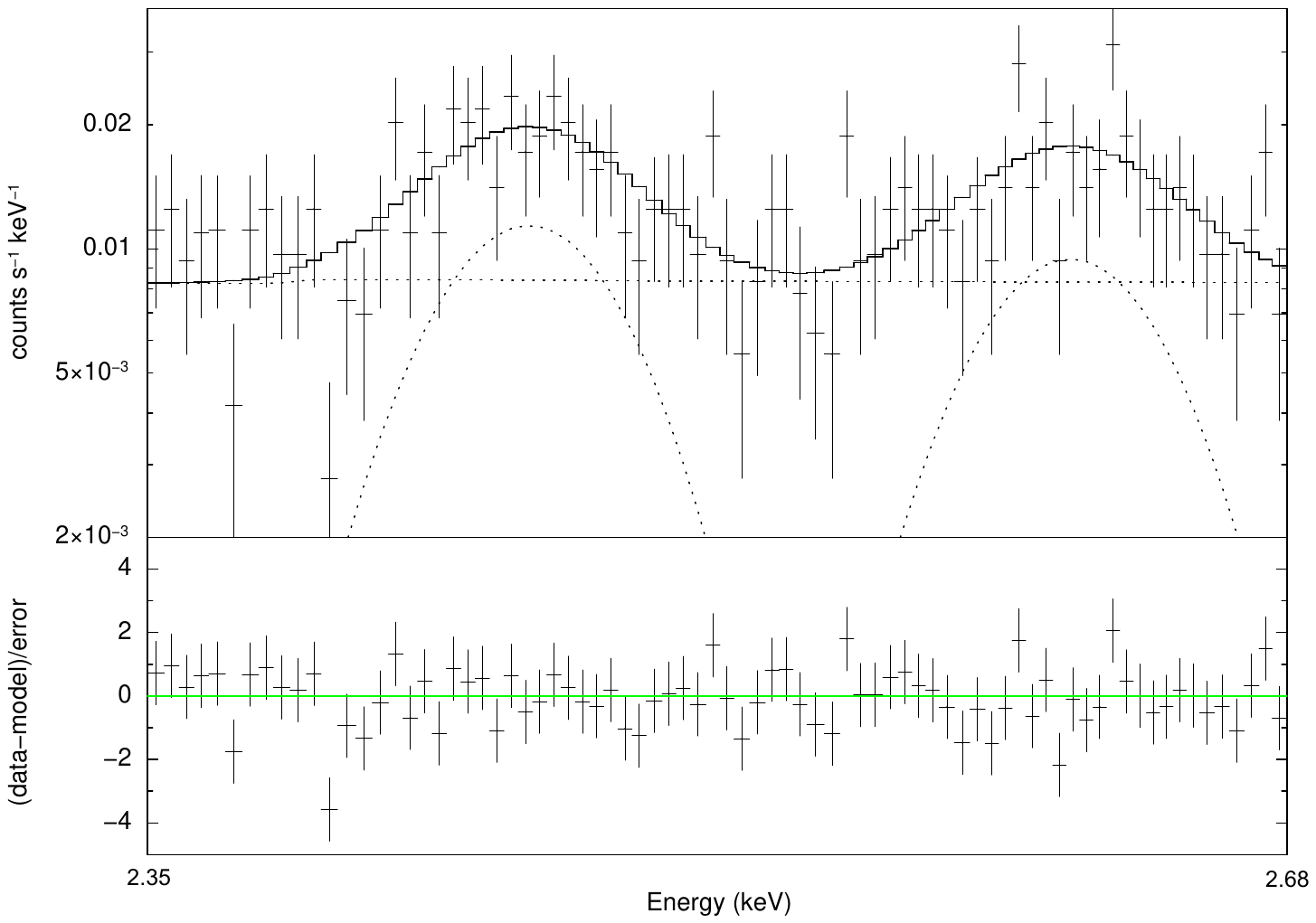}
    \includegraphics[trim={14pt 5pt 50pt 25pt},clip,width=\columnwidth]{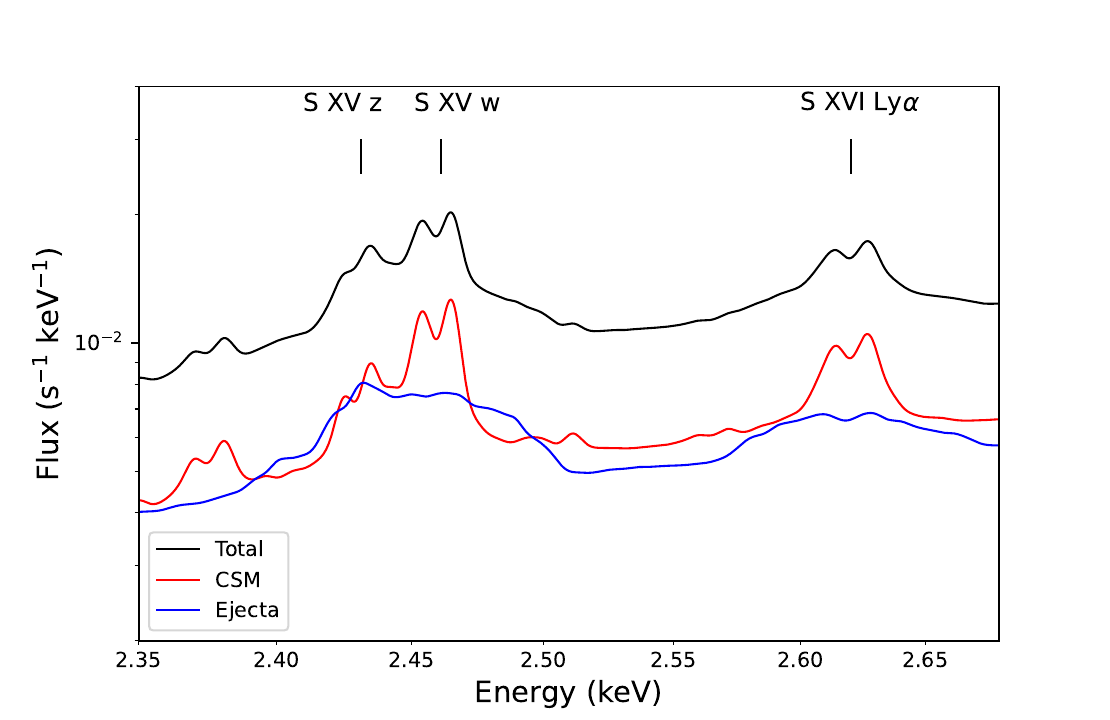}
    \caption{\textit{Upper panel:}Synthetic XRISM-Resolve spectrum of SN 1987A for the year 2024 (assuming the gate valve closed) with an exposure time of 160 ks including bulk motion Doppler broadening and binned using the \cite{2016A&A...587A.151K} optimal binning algorithm. 
    The synthetic XRISM-Resolve spectral model is superimposed with a black dashed line. 
    The red curve shows the contribution of the CSM (ER and \hii) to the spectral model, while the blue curve shows the contribution of the ejecta.
    \textit{Lower left panel:} Close-up view of the spectrum shown in upper panel in the $2.38-2.65$ keV band with best-fit model and residuals.
    \textit{Lower right panel:} Close-up view of the spectral model shown in upper panel in the $2.38-2.65$ keV band.
    }
    \label{fig:gvcspec}
\end{figure*}
Upper panel of Figure \ref{fig:gvcspec} shows the updated version of the synthetic spectrum, for an exposure time of 160 ks, which is the amended exposure time for the PV phase observation of SN 1987A, increased with respect to the original 100 ks to compensate the loss of effective area.
Unfortunately, the effective area below 2 keV is severely compromised, hampering us to carry on the diagnostic procedure for the Mg and Si emission lines.
However, the quality of the spectrum remains adequate for applying the same diagnostics to the S lines.
In particular, it will be possible to reveal the ejecta through their enhanced line widths (with respect to the narrow lines stemming from the shocked ambient medium) associated with their large velocities.
We then fitted the S line profiles adopting only 2 broad Gaussian components (due to the reduced statistic), for the S XV w-z and for the S XVI Ly$\alpha$, modeling the continuum emission with a power law and accounting for the interstellar and intergalactic absorption.
Similarly to \cite{2024ApJ...961L...9S} $\sigma_2= \sigma_1\cdot E_2/E_1$, being $\sigma_{1,2}$ and $E_{1,2}$ the width and the centroid of the ejecta components in the lines of He-like and H-like species.
Figure \ref{fig:gvcspec} (lower-left panel) shows a close-up view of the synthetic spectrum highlighting the S lines with the corresponding best fit model and residuals.
The different components are detected with high significance (their normalization being larger than zero at more than the 5$\sigma$ confidence level) and clearly show a large expansion velocity ($3300_{-700}^{+800}$ km s$^{-1}$), which nicely accounts for the ejecta bulk motion.
%\pagebreak
\section{Discussion and Conclusions}\label{sect:con}
We synthesized the updated PV XRISM-Resolve X-ray spectrum for 2024, including the effects of the gate valve closed. Our analysis shows that the synthetic spectra still provides valuable diagnostics to measure the ejecta velocity from the broadening of sulfur emission lines, highlighting the continued utility of XRISM in studying SN 1987A.

Our results showcase the adaptability and resilience of our analytical methods and the importance of ongoing X-ray observations and simulations in unveiling the complex dynamics of supernova remnants. Future observations, leveraging both current and upcoming missions, are expected to provide deeper insights into the evolving nature of SN 1987A.
%\vspace{8pt}

%{\noindent \small \textit{Acknowledgments} - This work was supported by JSPS Core-to-Core Program, (grant number:JPJSCCA20220002).
%This work was financially supported by Japan Society for the Promotion of Science Grants-in-Aid for Scientific Research (KAKENHI) Grant Number,  JP23H01211 (AB), JP20H00174 (S.K.) and JP21H01121 (S.K. and Y.T.).
%%MM, SO, and FB acknowledge financial contribution from the PRIN MUR "Life, death and after-death of massive stars: reconstructing the path from the pre-supernova evolution to the supernova remnant" and the Astrofund Theory Grant of INAF.
%S.N. was supported by JSPS KAKENHI (A) Grant Number JP19H00693 and RIKEN Pioneering Project for Evolution of Matter in the Universe (r-EMU).} 

\bibliography{sample631}{}
\bibliographystyle{aasjournal}
\end{document}